\documentclass{article}
\usepackage[final]{neurips_2025}
\makeatletter
\providecommand{\@trackname}{}
\makeatother
\usepackage{graphicx} 
\usepackage{outlines}

\usepackage[utf8]{inputenc} 
\usepackage[T1]{fontenc}    
\usepackage{hyperref}       
\usepackage{url}            
\usepackage{booktabs}       
\usepackage{amsfonts}       
\usepackage{nicefrac}       
\usepackage{comment}        
\usepackage{microtype}      
\usepackage{xcolor}         
\usepackage{comments}
\usepackage{float}
\usepackage{enumitem}

\title{Workflows vs Agents for Code Translation}
\workshoptitle{Deep Learning for Code in the Agentic Era}
\date{July 2025}
\author{
  Henry Gray \quad Tom Yotam \quad Octavian Udrea \\ \\
  Code Metal \\ 
  \texttt{\{henry.gray, tom, octavian\}@codemetal.ai
}
}
\begin{document}

\maketitle

\begin{abstract} \label{Abstract}
Translating algorithms from high-level languages like MATLAB to hardware description languages (HDLs) is a resource-intensive but necessary step for deployment on FPGAs and ASICs. While large language models (LLMs) offer a path to automation, their limited training on HDL code makes end-to-end transpilation brittle and prone to syntax errors. We compare two LLM-driven methods for syntax repair in a MATLAB-to-HDL pipeline: a structured, expert-designed flow that follows a fixed sequence of operations, and a more autonomous agentic approach that uses the Model Context Protocol (MCP) \cite{anthropic2024mcp} to dynamically select its own tools. We study 42 MATLAB signal-processing functions and isolate the syntax-repair stage. Across three model scales, the agentic approach is more effective at resolving initial syntax errors, unblocking a greater number of candidates to proceed through the pipeline. This upstream improvement yields measurable downstream improvements, most notably on mid-sized models, where it increases the simulation reach rate by over 20 percentage points. We hypothesize that the gains come from short prompts, aggressive context management, and conditional tool use. Conditional retrieval helps at 8B and 30B; at 235B final-success gains are small and a naive RAG variant attains the highest final success. Our findings suggest that these agentic frameworks, when properly designed, are most effective at compensating for the capacity limits of small and mid-sized models.
\end{abstract}

\section{Introduction} \label{Introduction}

Most digital signal processing algorithms are at least initially developed in MATLAB because it offers rapid iteration, a rich set of operations and toolboxes, and convenient testbench generation. However, deployment targets are often FPGAs or ASICs that demand code written in a hardware description language for low latency, high throughput, tight power budgets, and greater resource control. Manually bridging the gap between these paradigms, from MATLAB's high-level, dynamically typed environment to the low-level, statically typed structure of an HDL, is known to be slow and error-prone. Traditional compiler approaches, which are rule-based, are limited in their application to a subset of operations and rigid in the code format they generate. 

Although large language models (LLMs) promise to accelerate this process, their proficiency with HDLs is limited by the scarcity of high-quality open-source training data. As a result, direct end-to-end translation often fails, requiring a structured pipeline with robust guardrails. In such pipelines, failures frequently occur in the initial syntax repair stage. Naive, automated fixes at this step can silently alter the code's semantics, ultimately leading to verification failures.

In this paper, we focus on this critical syntax repair stage and conduct an empirical comparison of two LLM-driven strategies. We evaluated an expert-designed flow that follows a fixed script against a flexible agentic framework that uses the Model Context Protocol (MCP) to dynamically select tools. We study 42 MATLAB signal-processing functions and isolate the syntax-repair stage.

This paper makes three primary contributions. First, we provide a detailed empirical comparison of expert-designed and agentic flows across three model scales, quantifying their impact on pipeline success. Second, from these results, we derive a set of practical design guidelines for building effective agentic frameworks, emphasizing the importance of minimal prompts and aggressive context management. Finally, we isolate and demonstrate a key retrieval strategy, showing that on-demand tool use helps at small to mid scale, but naive inclusion of the same tools is actively detrimental to performance. 

Beyond the MATLAB to HDL setting, these results expose a broader pattern in agentic code systems. The experiments suggest that performance depends less on the presence of tools than on how and when they are invoked. Conditional retrieval, prompt minimalism, and separation of planning from generation emerge as scale-sensitive design levers that generalize to other agentic workflows such as refactoring, type inference, and translation between low-resource languages. We therefore view syntax repair as a controlled case study that reveals general principles for reliable tool use in language-model-driven programming.

\section{Related Work} \label{sec:related_works}

Our work is situated at the intersection of several key research areas: LLMs for code translation, agentic frameworks with tool use, and the specific challenge of translating code to hardware description languages (HDLs).

The use of large language models for programming tasks was measurably advanced by models like Codex, which demonstrated a strong ability to generate code in common languages \cite{chen2021evaluating}. However, these models often struggle with the ``long tail'' of specialized or low-resource languages. Recent work by Vijayaraghavan et al. (2024) specifically highlights this, showing that while LLMs can generate functionally correct VHDL, they often produce non-synthesizable or inefficient code that requires major manual correction \cite{vijayaraghavan2024vhdlEval}. This motivates our focus on a structured pipeline with a dedicated repair stage, rather than relying on direct, end-to-end translation.

Several recent efforts have focused specifically on the MATLAB-to-HDL problem. For instance, Schwartz et al. (2024) introduced a fine-tuning approach to improve an LLM's proficiency in VHDL, demonstrating gains but also noting the high cost of data acquisition \cite{schwartz2025}. Concurrently, Thakur et al. (2023) developed VeriGen, a system that uses an LLM to translate Python to Verilog for specific dataflow applications, relying on formal methods to constrain the output \cite{thakur2023verigen}. Our work complements these approaches by focusing not on model fine-tuning or formal constraints, but on a more flexible, agentic repair process that can be applied to general-purpose LLM outputs.

To overcome the limitations of standalone LLMs, recent research has focused on agentic frameworks that allow models to use external tools. Foundational work like ReAct \cite{yao2022react} established the core ``reason-act'' loop, while Toolformer \cite{schick2023toolformer} showed how models could learn to use APIs. Our MCP-based flow builds directly on these concepts, providing the LLM with a set of specialized tools (a compiler, a retrieval system) to diagnose and fix errors. A key contribution of our paper is the analysis of the \textit{strategy} for tool use. While the use of Retrieval-Augmented Generation (RAG) \cite{lewis2020rag} is a common technique, we demonstrate that for code repair, a conditional invocation policy is critical to its success.

\section{Experimental Setup} \label{Experimental Setup}

To evaluate the effectiveness of an agentic framework in a real-world programming task, we integrated two distinct syntax repair methodologies into a MATLAB-to-HDL transpilation pipeline\footnote{Transpilation commonly stands for translating compilation, a form of language to language translation.}. This setup allowed us to compare not only their immediate success in repairing syntax but also their impact on downstream verification and final code quality.

\subsection{Transpilation Pipeline} \label{Transpilation Pipeline}

\begin{figure}[H] 
    \centering
    \includegraphics[width=1\linewidth]{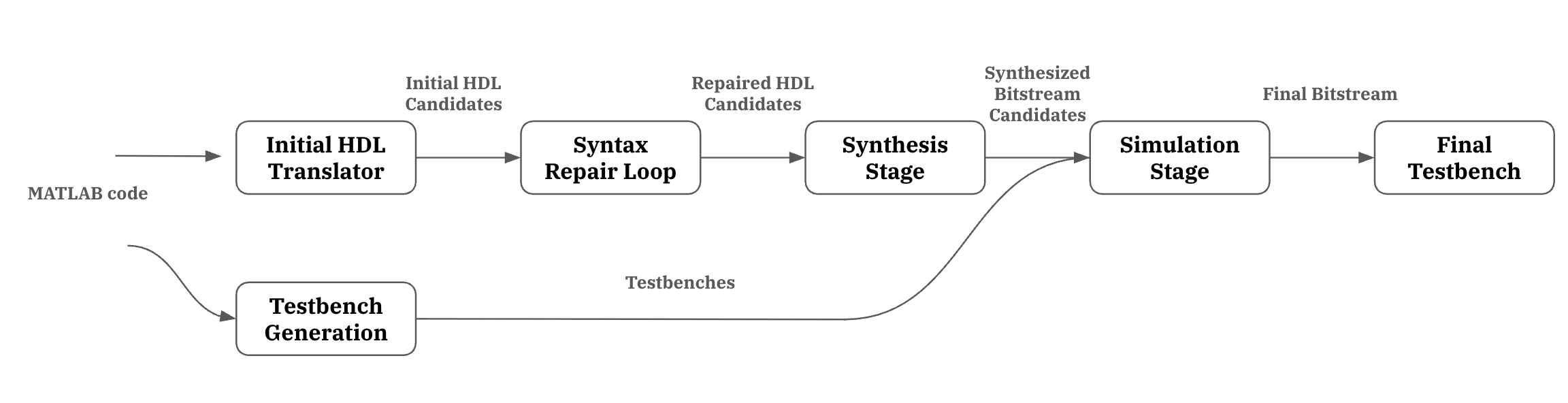}  
    \caption{MATLAB to HDL, end-to-end pipeline}
    \label{fig:baseline-flow}
\end{figure}

Our end-to-end pipeline, illustrated in Figure \ref{fig:baseline-flow}, begins by generating a testbench from the source MATLAB code that is used to verify the behavior of the HDL code generated. If necessary, the MATLAB code is then converted to fixed-point arithmetic and to handle data in a streaming rather than batch fashion. An LLM uses this version to generate multiple candidate HDL translations (in our case, 3). Because this initial translation frequently introduces errors, each candidate is sent to a dedicated syntax repair stage. This is the stage that we isolate for our experiment. After repair, syntactically correct code proceeds to a synthesis stage and is finally validated against the original testbench.

\subsection{Comparing Syntax Repair Flows} \label{Comparing Syntax Repair Flows}

\begin{figure}[H] 
    \centering
    \includegraphics[width=0.7\linewidth]{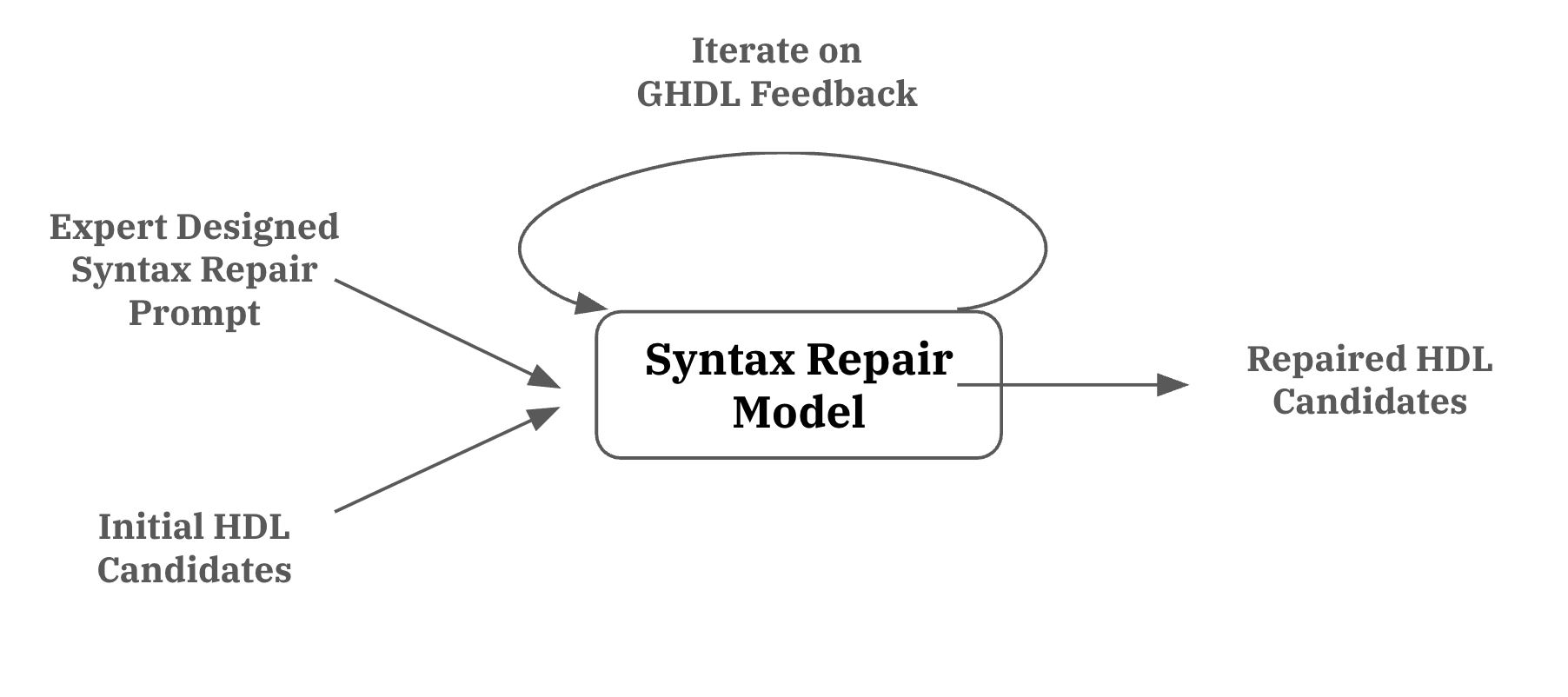}  
    \caption{The expert-designed baseline flow, which follows a fixed repair script.}
    \label{fig:baseline-repair}
\end{figure}

\textbf{Expert-Designed Flow (Baseline):} As shown in Figure \ref{fig:baseline-repair}, this flow represents a structured, non-agentic approach. For each HDL candidate, the LLM is provided with a large, expert-written prompt containing detailed guidance and advice for repairing HDL syntax. It uses the GHDL compiler to identify errors and iterates on the code in a fixed loop until syntax passes or a limit is reached.

\begin{figure}[H] 
    \centering
    \includegraphics[width=0.7\linewidth]{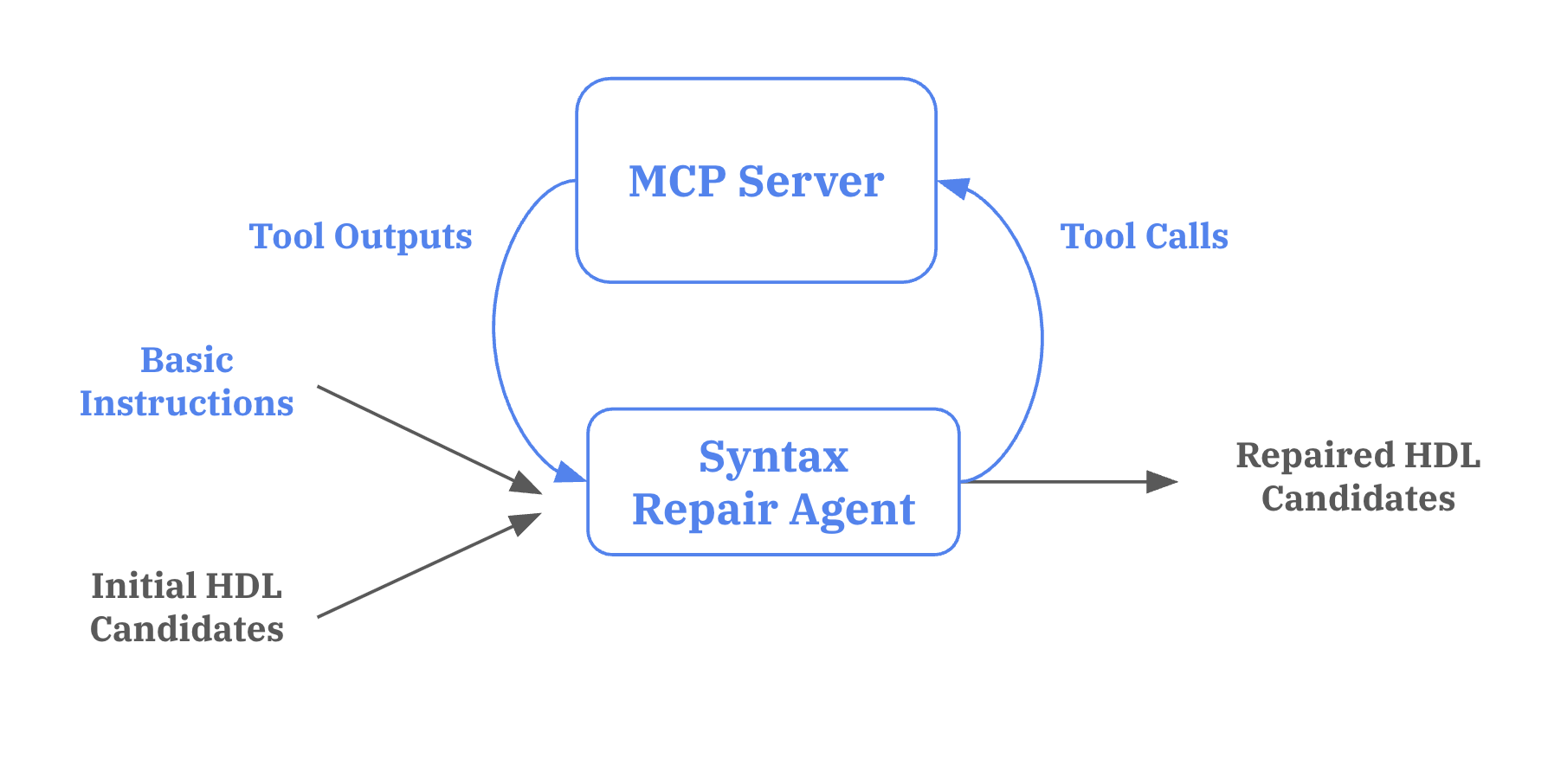}  
    \caption{The agentic MCP flow, where the agent dynamically selects tools in an iterative loop.}
    \label{fig:mcp-repair}
\end{figure}

\textbf{Agentic MCP Flow:} In contrast, the flow in Figure \ref{fig:mcp-repair} provides the LLM with autonomy. The agent receives a minimal prompt containing only the broken code, a repair goal, and a menu of available tools. It is then free to independently select and sequence tools to solve the problem. The tools provided were the following.

\begin{enumerate}
    \item \textbf{GHDL Syntax Check:} The same compiler used in the baseline flow to get a list of syntax errors.
    \item \textbf{RAG Retrieval:} A tool to retrieve syntactically correct VHDL code examples from a vector database to provide relevant context.
    \item \textbf{Code Rewrite:} A tool that passes the original code and a set of model-generated instructions to a second clean context agent for implementation. This design avoids context contamination from previous failed attempts.
\end{enumerate}

The context was aggressively pruned to maintain performance and avoid exceeding the context window of smaller models. This decision was motivated by preliminary experiments showing that including history from failed repair iterations decreased performance. Therefore, the context was reset after each attempt, with only a brief summary of the previous attempt carried over.

\subsection{Dataset and Evaluation Metrics} \label{Dataset and Evaluation Metrics}

We evaluated both flows on an internal dataset of 42 MATLAB functions for signal processing, ranging from 3 to 311 lines of code. We measured performance at both the candidate and function levels across three key metrics:
\begin{enumerate}
    \item \textbf{Syntax Pass Rate:} The percentage of candidates (or functions with at least one candidate) that successfully pass the GHDL syntax check.
    \item \textbf{Simulation Reach Rate:} The percentage of functions whose repaired code is valid enough to proceed to the synthesis and simulation stage.
    \item \textbf{Final Flow Success Rate:} The percentage of functions whose final HDL output passes the MATLAB-generated testbench, confirming semantic equivalence.
\end{enumerate}

\subsection{RAG Corpus and Retrieval Configuration} \label{sec:rag_config}
We built a retrieval corpus of syntactically correct VHDL functions drawn from internal examples used in class exercises and small design blocks. More than 1000 functions were included for reference by the tool. The functions were short, generally under 100 lines. Each item was stored as a single document representing a complete function, not fragmented into overlapping chunks. This whole function policy preserved idioms such as library imports, type declarations, and process structure that matter for synthesis.

We embedded each document using Infinity embeddings \cite{feil_2023_11630143}. At query time the agent formed a text query from the current error context and the working VHDL variant, then retrieved the top 3 nearest neighbors. The examples were inserted verbatim as full functions, subject to a strict token budget with truncation rules that favored keeping headers, library lines, and process scaffolding.

In the naive Non MCP with RAG variant, the top 3 examples were appended unconditionally to every repair prompt. In the MCP variant, retrieval was conditional on lack of progress or on compiler errors that indicated missing idioms, and the agent could choose to skip retrieval. No additional reranking by compiler error similarity was used.

We used three Qwen3 checkpoints from Hugging Face, Qwen3-8B \cite{qwen3-8b}, Qwen3-30B-A3B \cite{qwen3-30b-a3b}, and Qwen3-235B-A22B \cite{qwen3-235b-a22b}, with default tokenizers and no additional fine-tuning.

\subsection{Reproducibility.}
Additional configuration and decoding details are provided in Appendix~\ref{appendix:reproducibility}.

\section{Experimental Results} \label{Experimental Results} 
We evaluated the expert-designed (non-MCP) and agentic (MCP) syntax repair flows on our 42-function MATLAB dataset using Qwen across three model scales: 8B, 30B, and 235B. The results demonstrate that the agentic MCP approach consistently improves pipeline progression, with the most measurable impact observed at the mid-scale 30B model. 

Variability across twelve independent runs was small ($\sigma \approx$ 2-3 pp on intermediate metrics in pilot logging).

\subsection{MCP Measurably Improves Pipeline Progression at 30B} At the 30B scale (Table~\ref{tab:30b_headline}), the agentic MCP approach yields substantial improvements over the expert-designed baseline across all intermediate metrics. The candidate-level syntax pass rate increases from 51.9\% to 75.0\% (+23.1 pp), and the function-level syntax pass rate increases from 81.2\% to 92.3\% (+11.1 pp). The most dramatic gain is in the share of functions that reach the simulation stage, which jumps from 72.1\% to 95.3\% (+23.2 pp). This upstream success translates into a measurable improvement in the end-to-end success rate, which improves from 33.5\% to 42.1\% (+8.6 pp). These results indicate that while downstream semantic issues remain a bottleneck, MCP is highly effective at resolving the initial syntax errors that cause the most pipeline attrition. For reference, the naive variant Non-MCP + RAG underperforms at 30B; see Table~\ref{tab:30b_headline}, where the simulation reach is 44.0\% and the final success is 19.5\%.

\subsection{The Effect of MCP is Scale-Dependent} The benefits of the agentic framework vary with the size of the model, as shown in Tables~\ref{tab:8b_headline} and \ref{tab:235b_headline}. 
\begin{itemize} 
\item \textbf{At 8B}, MCP provides a crucial lift for the smaller model, improving the function-level syntax pass from 76.7\% to 90.7\% (+14 pp) and boosting the simulation reach rate by over 20 pp. However, the model's limited capacity constrains its ability to translate these gains into end-to-end success, which sees a more modest improvement (+4.9 pp). 
\item \textbf{At 235B}, the baseline expert-designed flow is already highly competent, successfully repairing syntax for 93\% of functions. Here, MCP has less headroom to add value. It pushes the function-level syntax pass to 100\% and provides a small lift to the final success rate (+2.3 pp), but its overall impact is diminished. In particular, the naive variant non-MCP + RAG attains the highest final success at 235B (58.1\% vs 55.8\% for MCP; Table~\ref{tab:235b_headline}); we defer the analysis to Section~\ref{Discussion}.

\end{itemize} 

\subsection{Conditional Tool Use is Critical; Naive RAG Inclusion Degrades Performance} To isolate the impact of tool-use policy, we tested a variant that naively appended RAG outputs to every repair prompt. As shown in the Non-MCP+RAG columns of Tables~\ref{tab:8b_headline}, \ref{tab:30b_headline}, and \ref{tab:235b_headline}, unconditional inclusion is detrimental at smaller and mid scales.
 At the 30B scale (Table~\ref{tab:30b_headline}), naively adding RAG caused the simulation reach rate to drop from 72.1\% to 44.0\% and the final success rate to drop from 33.5\% to 19.5\%.
 This provides strong evidence that the agentic framework's success is driven not just by the availability of tools, but by its ability to apply them conditionally and avoid the context clutter that harms less capable models.

\begin{table}[H]
  \centering
  \caption{Qwen 30B, function-level macro averages (baseline, MCP, and baseline + naive RAG).}
  \label{tab:30b_headline}
  \begin{tabular}{lccc}
    \toprule
    Metric & Non-MCP & MCP & Non-MCP+RAG \\
    \midrule
    Candidate-level syntax pass & 51.9\% & \textbf{75.0\%} & 60.0\% \\
    Function-level syntax pass   & 81.2\% & \textbf{92.3\%} & 77.0\% \\
    Reach testbench              & 72.1\% & \textbf{95.3\%} & 44.0\% \\
    Final success                & 33.53\% & \textbf{42.12\%} & 19.5\% \\
    \bottomrule
  \end{tabular}
\end{table}

\begin{table}[H]
  \centering
  \caption{Qwen 8B, function-level macro averages (baseline, MCP, and baseline + naive RAG).}
  \label{tab:8b_headline}
  \begin{tabular}{lccc}
    \toprule
    Metric & Non-MCP & MCP & Non-MCP+RAG \\
    \midrule
    Candidate-level syntax pass & 59.0\% & \textbf{63.1\%} & 56.7\% \\
    Function-level syntax pass   & 76.7\% & \textbf{90.7\%} & 76.7\% \\
    Reach testbench              & 60.5\% & \textbf{90.7\%} & 60.5\% \\
    Final success                & 18.3\% & \textbf{23.2\%} & 16.9\% \\
    \bottomrule
  \end{tabular}
\end{table}

\begin{table}[H]
  \centering
  \caption{Qwen 235B, function-level macro averages (baseline, MCP, and baseline + naive RAG).}
  \label{tab:235b_headline}
  \begin{tabular}{lccc}
    \toprule
    Metric & Non-MCP & MCP & Non-MCP+RAG \\
    \midrule
    Candidate-level syntax pass & 86.0\% & \textbf{94.4\%} & 93.9\% \\
    Function-level syntax pass   & 93.0\% & \textbf{100\%}  & 100\% \\
    Reach testbench              & 100\%  & 100\%           & 100\% \\
    Final success                & 53.5\% & 55.8\%          & \textbf{58.1\%} \\
    \bottomrule
  \end{tabular}
\end{table}

\begin{figure}[H] 
    \centering
    \includegraphics[width=0.7\linewidth]{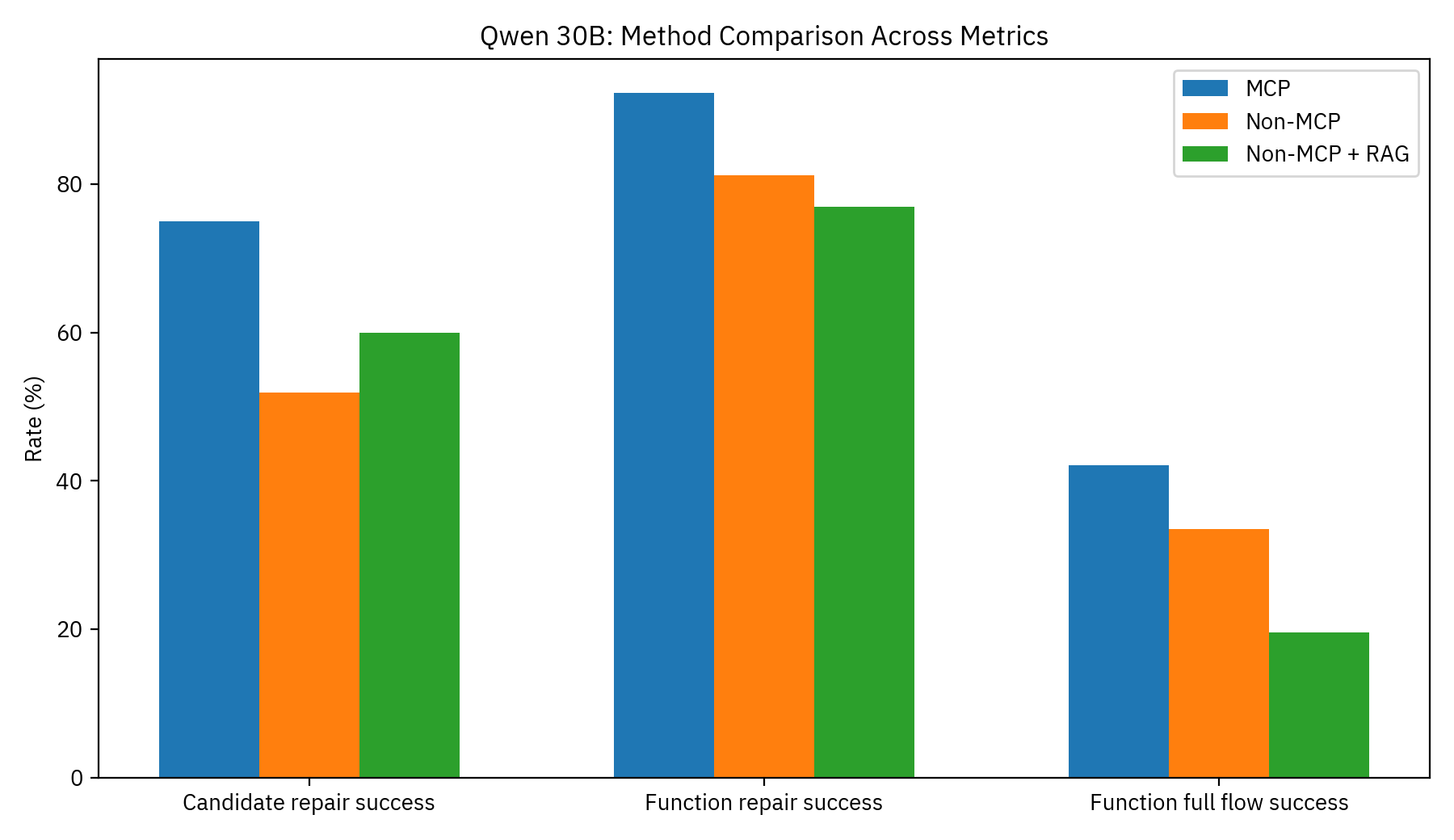}  
    \caption{metrics comparison for 30B model}
    \label{fig:mcp-flow}
\end{figure}

\section{Discussion} 
\label{Discussion}
\subsection{Conditional Tool Use is the Primary Driver of Improvement}
Our results show that an agentic framework using MCP for syntax repair measurably improves pipeline progression compared to a deterministic, expert-designed flow. The success of this approach, however, depends heavily on a set of core design principles and the underlying model's scale.

To understand why MCP helps, we consider how the framework structures reasoning within the context window. We hypothesize that these improvements arise from differences in prompt entropy and token allocation rather than from the tools themselves. When the model must process long, unfiltered retrieval outputs, its effective reasoning bandwidth is diluted. Conditional invocation instead maintains a compact, high-signal context that fits within the model’s limited working memory. While we did not directly test this mechanism, the scale pattern in Tables 1-3 is consistent with this interpretation.

The central finding of our work is that how auxiliary information is introduced matters as much as what information is available. At the 30B scale, the agentic approach delivered large improvements to intermediate metrics, most notably lifting the simulation reach rate from 72.1\% to 95.3\%. This gain is directly attributable to a conditional tool-use policy.

This policy was encoded in a prompt that created a simple "if stuck, then retrieve" loop, instructing the agent to first attempt a local fix and only invoke RAG when it failed to make progress. We found that RAG helps when an error requires nonlocal structure or idioms absent from the current context, but it hurts when it injects loosely related code that widens the search space or distracts the model.

This was validated by our negative control experiment, where naively appending RAG outputs to the expert-designed flow degraded performance sharply. At 30B, this unconditional retrieval caused the final success rate to drop from 33.5\% to 19.5\%. We identified three failure modes for this approach: context clutter from large examples, architectural mismatches in retrieved code, and the truncation of precise compiler errors. The agentic framework's ability to selectively deploy tools avoids these pitfalls and is therefore its key advantage.

This pattern suggests that agentic frameworks contribute value not by adding new capabilities, but by enforcing structure that protects limited attention within the model’s context window. This observation aligns with prior findings on reasoning bandwidth limits in LLMs and suggests that agentic orchestration may act as a form of context regularization.

\subsection{The Interplay of Agentic Design and Model Scale}
The effectiveness of the MCP framework is clearly dependent on the model's intrinsic capacity. We observed two forces interacting with scale: the planning and selectivity which MCP supplies, and the intrinsic capacity which depends on the model's size.

When capacity is low (8B), MCP's structure enables progress but cannot fully compensate for limited semantic modeling. It provides a crucial boost to pipeline progression but only a modest lift to final success.

When capacity is moderate (30B), MCP's selective tool use and context hygiene are complementary, yielding the greatest overall uplift. The model has enough reasoning bandwidth to exploit the framework's design, converting a large number of otherwise failing candidates into successful simulations.

When capacity is high (235B), the model's intrinsic ability saturates progression metrics, leaving only a narrow semantic frontier where MCP can help. The framework's benefit is real but small relative to its overhead.

This pattern suggests that agentic frameworks are most impactful at the mid-scale, where they effectively compensate for a model's capacity limits without being rendered redundant by its sheer competence.

These findings indicate that policy of tool use is a controllable design variable. Conditional, selective invocation yields greater benefit at smaller model scales, whereas naive inclusion suffices or even helps at extreme scale. The practical takeaway is that reliable agentic systems depend as much on disciplined orchestration as on model capacity. Although the present study isolates the syntax-repair stage, the same orchestration principles-conditional tool use and separation of planning from generation likely extend to broader software-engineering workflows where LLMs operate under token or attention constraints.

\subsection{Naive RAG at 235B}
At 235B, the naive Non MCP with RAG variant attains the highest end to end success (58.1\% versus 55.8\% for MCP; Table~\ref{tab:235b_headline}). Intermediate metrics are saturated at this scale. Function level syntax pass is 100\% for both variants, and reach to testbench is 100\% for all methods. The gain therefore most likely arises at the final verification stage through canonicalization rather than earlier unblocking.

A simple hypothesis explains the pattern. Larger models filter non helpful retrieved tokens and are less susceptible to context saturation. Naive retrieval then acts as few shot priming instead of a distraction. In smaller and mid scale models, MCP helps by keeping prompts short and focused. At 235B that focusing benefit is largely endogenous to the model.

The practical implication is a hybrid policy for large models. Attach a compact deduplicated top k exemplar set that persists across attempts, and keep MCP for compiler guided diagnosis. Enforce strict length control to preserve signal to noise. The observed advantage is small at 2.3 percentage points, so it may fall within run to run variance.

We treat this as a testable claim. Run length controlled ablations with equal token budgets, inject noise into retrieval to probe filtering, rerank retrieval by compiler error similarity, disable MCP resets to test persistence effects, and report confidence intervals to establish whether the gain is statistically reliable.

These numbers (Tables~\ref{tab:8b_headline}, \ref{tab:30b_headline}, \ref{tab:235b_headline}) summarize the scale pattern: naive Non\textendash MCP+RAG harms 8B and 30B but attains the highest final success at 235B; we treat the filtering and context\textendash saturation explanation as a hypothesis pending length\textendash controlled ablations and CIs.

\subsection{Design evolution and lessons} \label{sec:design_evolution}
Our design converged through four stages. The first prototype reused the prior framework that received the full VHDL and was instructed to output new VHDL in \texttt{<vhdl>} tags. We added a small tool set: a manual lookup, a similar examples retrieval tool, and a helper that proposed subproblems. The agent called tools directly, accumulated all tool outputs in the prompt, and deferred code generation. This version rarely produced code before the context filled. Prompt tuning reduced tool calls, but performance still lagged the expert script.

The second stage allowed the agent to interleave thinking and tool calls and to place tool outputs into a persistent reasoning trace, then generate. Syntax pass improved, but context length again became the bottleneck.

The third stage switched to a suggest change policy. The agent proposed edits at specific line numbers, with the latest VHDL always shown in context. We added visible line numbers and required each edit to name the lines and include a tight local window around the change. This reduced tokens and worked on small refactors, but it failed when many declarations or process blocks needed coordinated changes.

The final stage separated analysis and generation. A tool using planner produced a compact instruction list and short rationale in a clean format. A second generator, with a clean context, produced a complete new VHDL unit from those instructions. This separation increased candidate level syntax pass and function level reach across scales. It also made behavior more stable, since the generator never saw long tool transcripts or prior failed attempts.

Tool ablations informed the final menu. The manual lookup tool reduced performance whenever it was invoked, so we removed it. The similar examples retrieval tool helped within MCP when used conditionally and under a strict token budget, but naive attachment to every prompt reduced performance at 8B and 30B while slightly helping at 235B. This pattern supports a policy conclusion rather than a tool conclusion: conditional use matters more than tool availability.

Why this worked: separating planning from generation reduced context contamination and kept token budgets predictable; short instruction payloads preserved signal to noise; and resets avoided anchoring on earlier failed edits. These mechanics align with the observed scale effects. At small and mid scale, MCP limits distraction and improves reach. At large scale, model capacity filters non helpful tokens so naive retrieval can act as few shot priming, which explains the small end to end gain at 235B.

For reproduction, report the corpus construction in Section~\ref{sec:rag_config}, the embedding model identifier, the top k used for retrieval, any deduplication, and fixed token budgets for planner instructions and inserted examples. Equalize token budgets when comparing policies to separate policy effects from length effects.

\subsection{Limitations and Future Work}
Our dataset consists of 42 functions from a single domain (signal processing), so these results may differ with other error profiles. The primary improvements we observed were in preventing pipeline attrition before the final verification stage, which indicates that a measurable semantic bottleneck remains downstream of syntax repair. Future work should focus on pairing agentic repair with semantic safeguards, such as differential testing or lightweight equivalence checks, to ensure that the gains from improved syntax repair fully propagate to the end-to-end success rate. 

The dataset used in this study is currently internal to our company. We plan to release a cleaned subset of the 42 MATLAB functions and corresponding HDL outputs in the near future to support external replication and follow-up work.

\section*{Broader Impacts}
This work studies agentic tool-use for program repair and translation. Although evaluated on MATLAB$\rightarrow$HDL, the design lessons (conditional retrieval, context control, separation of planning and generation) apply to agentic programming workflows such as refactoring, porting, linting, and large-scale code health tasks. Potential benefits include faster prototyping, lower entry barriers for hardware and systems development, and safer automation compared to unconstrained prompting.

Risks include plausible-but-wrong repairs that can introduce latent defects or security vulnerabilities; propagation of license-incompatible or proprietary snippets via retrieval; over-reliance on non-deterministic systems in safety- or mission-critical settings; and increased energy use from large models. Agentic systems can also widen the attack surface (e.g., prompt or retrieval injection) and may amplify biases present in code corpora.

Mitigations we recommend: enforce verification gates (compilation, unit and differential tests, fuzzing, and lightweight formal checks where feasible) before deployment; use privacy-preserving, in-tenant retrieval with provenance and license scanning; prefer conditional tool invocation and short prompts to reduce context leakage; log all tool calls and code diffs for audit; restrict side-effecting tools and require explicit human approval for high-impact actions; document stochasticity and provide reproducible decoding settings for review. The results here should be viewed as improving reliability in early pipeline stages, not as a substitute for semantic validation or secure development practices.

\bibliographystyle{plain}
\bibliography{refs.bib}

@article{chen2021evaluating,
  title   = {Evaluating Large Language Models Trained on Code},
  author  = {Mark Chen and Jerry Tworek and Heewoo Jun and Qiming Yuan and Henrique Ponde de Oliveira Pinto and Jared Kaplan and Harri Edwards and Yuri Burda and Nicholas Joseph and Greg Brockman and Alex Ray and Raul Puri and Gretchen Krueger and Michael Petrov and Heidy Khlaaf and Girish Sastry and Pamela Mishkin and Brooke Chan and Scott Gray and Nick Ryder and Mikhail Pavlov and Alethea Power and Lukasz Kaiser and Mohammad Bavarian and Clemens Winter and Philippe Tillet and Felipe Petroski Such and Dave Cummings and Matthias Plappert and Fotios Chantzis and Elizabeth Barnes and Ariel Herbert-Voss and William Guss and Alex Nichol and Alex Paino and Nikolas Tezak and Jie Tang and Igor Babuschkin and Suchir Balaji and Shantanu Jain and William Saunders and Christopher Hesse and Andrew N. Carr and Jan Leike and Josh Tobin and Jakub Pachocki and Aitor Ormazabal and Bob McGrew and Ilya Sutskever and Wojciech Zaremba},
  journal = {arXiv preprint arXiv:2107.03374},
  year    = {2021}
}

@article{schwartz2025,
  title   = {Fine-Tuning a Foundational LLM for VHDL Code Generation},
  author  = {Joshua Schwartz and Matthew J. Kusner and T. J. O'Donnell},
  journal = {arXiv preprint arXiv:2405.09610},
  year    = {2024}
}

@misc{qwen3-8b,
  title        = {Qwen3-8B},
  author       = {{Qwen Team}},
  year         = {2025},
  howpublished = {\url{https://huggingface.co/Qwen/Qwen3-8B}},
  note         = {Accessed 2025-08-27}
}

@misc{qwen3-30b-a3b,
  title        = {Qwen3-30B-A3B},
  author       = {{Qwen Team}},
  year         = {2025},
  howpublished = {\url{https://huggingface.co/Qwen/Qwen3-30B-A3B}},
  note         = {Accessed 2025-08-27}
}

@software{feil_2023_11630143,
  author    = {Feil, Michael},
  title     = {Infinity - To Embeddings and Beyond},
  year      = {2023},
  month     = {oct},
  publisher = {Zenodo},
  doi       = {10.5281/zenodo.11630143},
  url       = {https://doi.org/10.5281/zenodo.11630143}
}

@inproceedings{yao2022react,
  title={ReAct: Synergizing Reasoning and Acting in Language Models},
  author={Yao, Shunyu and others},
  booktitle={ICLR},
  year={2023},
  url={https://arxiv.org/abs/2210.03629}
}

@inproceedings{schick2023toolformer,
  title={Toolformer: Language Models Can Teach Themselves to Use Tools},
  author={Schick, Timo and others},
  booktitle={NeurIPS},
  year={2023},
  url={https://openreview.net/pdf?id=Yacmpz84TH}
}

@article{lewis2020rag,
  title={Retrieval-Augmented Generation for Knowledge-Intensive NLP Tasks},
  author={Lewis, Patrick and others},
  journal={arXiv:2005.11401},
  year={2020}
}

@misc{anthropic2024mcp,
  title={Introducing the Model Context Protocol},
  author={{Anthropic}},
  year={2024},
  url={https://www.anthropic.com/news/model-context-protocol}
}

@article{thakur2023verigen,
  title={VeriGen: A Large Language Model for Verilog Code Generation},
  author={Thakur, Shailja and others},
  journal={arXiv:2308.00708},
  year={2023},
  doi={10.1145/3643681}
}

@inproceedings{vijayaraghavan2024vhdlEval,
  title={{VHDL-Eval}: A Framework for Evaluating Large Language Models in VHDL Code Generation},
  author={Vijayaraghavan, Prashanth and others},
  booktitle={LAD 2024},
  year={2024},
  url={https://arxiv.org/abs/2406.04379}
}

@misc{qwen3-235b-a22b,
  title={Qwen3-235B-A22B},
  author={{Qwen Team}},
  year={2025},
  howpublished={HuggingFace model card},
  url={https://huggingface.co/Qwen/Qwen3-235B-A22B}
}

\appendix








\section{Reproducibility Details}
\label{appendix:reproducibility}

Models: Qwen3-8B, Qwen3-30B-A3B, Qwen3-235B-A22B (default tokenizers). 
Decoding: temperature=0.6, top-p=1.0, top-k unset, max-new-tokens backend default (unset), no stop tokens. 
Trials: R=12 independent runs per function; K=3 candidates per run. 
Repair loop: max iterations T=10; stop on first syntax pass or when T is reached. 
Retrieval: Infinity embeddings; $\ell_2$-normalize, FAISS IndexFlatIP (cosine), top-$k$=3; truncate retrieved examples to 1200 tokens keeping headers/imports/process scaffolding; retrieval is conditional on no-progress or GHDL errors matching \texttt{missing library/use/type/port/process}. 
Tooling: GHDL with \texttt{--std=08}; backend via Hugging Face endpoints or local Transformers $\ge$4.43 (insensitive under fixed decoding). 
Metrics: per-function candidate pass $p_i=\#$passes$/ (R\!\times\!K)$; function pass, reach, final success in $\{0,1\}$; report macro means over 42 functions. 

\subsection{Compute resources}\label{sec:compute}
Provider: Hugging Face Inference Endpoints. Backend: TGI (default image; version not recorded).

\textbf{Qwen3-8B.}
Instance: 1× NVIDIA A100 80GB GPU; 11 vCPUs; 145 GB RAM. Autoscaling: min 0, max 1 replica; strategy “hardware usage”; idle scale-to-0 after 15 min. Concurrency: single replica.
Runtime: total 4{,}933 min across 42 functions × R=12 trials (K=3, T=10), i.e., $\approx$82.2 GPU-hours.
Per-trial wall-clock (one function, K=3, T=10): $\approx$9.79 min.
Tokens: no explicit max-token limits set.

\textbf{Qwen3-30B.}
Instance: 1× NVIDIA H200 141GB GPU; 23 vCPUs; 256 GB RAM. Autoscaling: min 0, max 1 replica. Concurrency: single replica.
Runtime: total 5{,}388 min across 42 functions × R=12 trials (K=3, T=10), i.e., $\approx$89.8 GPU-hours.
Per-function trial wall-clock (K=3, T=10): $\approx$10.69 min.
Tokens: no explicit max-token limits configured.

\textbf{Qwen3-235B.}
Environment: on-prem DGX (local). Backend: Transformers/TGI (version not recorded).
Hardware: not logged. Replication guidance:
\begin{itemize}[nosep,leftmargin=*]
\item FP16: $\ge$8$\times$ 80 GB GPUs (e.g., 8$\times$A100 80GB) with tensor parallelism $\ge$8.
\item INT8/4-bit AWQ: $\ge$4$\times$ 80 GB GPUs (e.g., 4$\times$A100 80GB) with tensor parallelism $\ge$4.
\item System RAM $\ge$512 GB; fast local storage for model weights; recent CUDA driver.
\end{itemize}
Autoscaling/concurrency: not applicable (single local server). Runtime and wall-clock: not logged; replication may expect slower throughput than 30B under the same decoding settings.

\end{document}